\documentclass[twocolumn,showpacs,preprintnumbers,amsmath,amssymb]{revtex4}
\usepackage{graphicx}
\usepackage{dcolumn}
\usepackage{bm}

\begin{document}

\title{Spin-phonon coupling, q-dependence of spin excitations 
and high-T$_C$ superconductivity from
band models.}

\author{T. Jarlborg}

\affiliation{
DPMC, University of Geneva, 24 Quai Ernest-Ansermet, CH-1211 Geneva 4,
Switzerland
}


\begin{abstract}

An understanding of spin excitations in cuprates is
essential since 
the mechanism of high-T$_C$ superconductivity might be linked to spin fluctuations.
Band calculations for long "1-dimensional" unit cells of La$_2$CuO$_4$ show that 
the coupling between antiferromagnetic spin waves
and phonons is larger for distortions of oxygens than for Cu or La.
When this result is applied to a 2-dimensional, free-electron like
band, it leads to a q-dependent spin
excitation spectrum in good agreement with recent experiments.
It is argued that important parameters for spin-phonon coupling,
which comes out from the comparison between experiment and theory,
are relevant for the mechanism of superconductivity, and are large enough to explain a high T$_C$.

\end{abstract}


\maketitle

Many measurements have revealed the existence of complex superstructures
and pseudogaps
in high-T$_C$ copper oxides.  A picture of spatial, stripe-like
regions with excitations of charge
and spin modulations emerged quite early from neutron scattering
experiments \cite{tran}.  
Angular resolved photoemission confirms that the bandstructure
agree well with calculations, where one dispersive band
gives rise to an ubiquitous barrel-like Fermi-surface (FS) (or a FS-"arc" when displayed
in 1/4th of the Brilloin zone) \cite{seppo,dama}. However, this arc is truncated
and survives only in the diagonal direction ($k_x \approx k_y$)
at low temperature, $T$, \cite{nor}. Phonons enter also
into this complexity, as shown by the softening of
some phonon branches at certain dopings \cite{uchi,fuk,pint},
by isotope effects on the pseudogap near the
temperature $T^*$, on T$_C$ \cite{rub}, and on the gap structure itself \cite{gweo}.
Non-commensurate, $q$-dependent spin excitations
appear as side-spots in neutron scattering at (0.5-$q$,0.5)
and (0.5,0.5-$q$), where $q$ varies linearly as function of doping, $x$,
up to a saturation at $x \approx 0.12$ \cite{yam}. Recent measurements have established that
the underlying spin excitations have a characteristic "hour-glass" shaped ($\bar{q},\omega$)-dispersion,
even at non-superconducting compositions \cite{hay,tran1,vig}.
Here, it is shown that the characteristic spin dispersion can be understood in terms
of spin-phonon coupling (SPC), and secondly it is concluded that
phonons promote equal spin pairing and a large $\lambda_{sf}$ for spin fluctuations,
and therefore SPC is important for the mechanism of superconductivity.

Previous ab-initio band calculations for supercells containing 1-dimensional (1D) phonon 
and spin-wave modulations in the CuO bond direction ([1,0,0]), 
show large SPC
within the CuO plane of these systems \cite{tj1}. 
This means that an antiferromagnetic (AFM) wave of the correct wave length and the proper phase
is stronger when it co-exists with the phonon \cite{tj3}. 
These results,
in combination with 2D free-electron like bands, have been used for modeling of many
normal state properties of the high-T$_C$ materials \cite{tj5,tj6}.
Phonon softening, dynamical stripes, correlation between $\bar{q}$ and $x$,
smearing of the non-diagonal part of the FS, and abrupt disappearance of the spin
fluctuations at a certain $T^*$, are possible consequencies of SPC within a 
rather conventional band. The present calculations consider SPC between spin
waves and four different types of phonon distortions.

First, we discuss the ab-initio band calculations made for
La$_{(2-x)}$Ba$_x$CuO$_4$ (LBCO), where the virtual crystal approximation (VCA) is applied
to La-sites to account for the doping.
The calculations are made for long supercells oriented along
the CuO bond direction, by using the linear Muffin-Tin Orbital method (LMTO) in
the local spin-density approximation (LSDA), as has been described previously \cite{tj5}.
Phonon distortion amplitudes ($u$) and the size of Cu moments ($m$) in spin waves
are necessary input to these calculations.
 The $T$-dependences 
$u^2 \approx 3k_BT/K_u$ and $m^2 \approx k_BT/K_m$ are valid for not too low $T$.
Here $K_p = d^2E/dp^2$, $E$ is the total energy, and $p=u$ or $m$, respectively \cite{tj5}.
The force contants $K_u$ for the different atoms 
are taken from measurements on YBa$_2$Cu$_3$O$_7$ \cite{hum,thom}. The resulting $u/a_0$,
where $a_0$ is the lattice constant, are shown in the Table for $T \approx 100K$
(the interesting temperature range for $T_c$ and $T^*$). 
An
approximate calculation of $K_m$ for a short wave in HgBa$_2$CuO$_4$ corresponds to
magnetic moments $m$ on Cu of about $\sim 0.09 \mu_B$ at 100 K \cite{tj3}. Such moments are typically obtained
by application of a magnetic field of the order of $\pm 5$ mRy \cite{tj5}.

Results of calculations are presented here showing a varying degree of SCP for different phonons.
These calculations consider 
distortions in a cell with 16 formula units (8$a_0$ along $\bar{x}$) 
with displacements of La and apical O
along $\bar{z}$, and Cu and planar-O along $\bar{x}$, and the phonons may co-exist with AFM spin-waves.
For the latter phonon there is a
positive SPC when the nodes of the AFM 
waves are located at the "compressed" Cu sites (when the O-atoms are moved towards the Cu). 
Optimal SPC for displacements
of out-of-plane atoms (La and apical O) occurs when these atoms move towards the CuO plane
near the region of the AFM node.
The calculated results for the maximal potential shifts on Cu caused by phonons ($V_q^p$) and by spin waves
($V_q^m$ in the spinpolarized potential) 
for the 4 types of movements in the cell are shown in the Table, together with
the result for $V_q^m$ without phonon.
That the phase between the phonon and spin wave is crucial for having a
positive SPC is shown by the fact that 
$V_q^m$ for an "apical-oxygen" phonon is reduced from 10 to below 8 mRy when the phase 
between the phonon and the spin wave is shifted by 
$\pi$.

Measured \cite{thom,hum} and calculated \cite{and,coh} phonon frequencies for zone boundary phonons for YBa$_2$Cu$_3$O$_7$
are indicative for typical $\hbar\omega$ for each atomic character of the phonon DOS. Phonon energies 
and partial phonon DOS for each site, $N_{site}$ have been calculated by Chen and Callaway \cite{chen}
for a similar cuprate system, Nd$_2$CuO$_4$, and the
same $N_{site}$ can be assumed to be representative for LBCO.  
From these references it is seen that the main La modes are 
at 10-20 meV, Cu at 20-30 meV, planar O near 50$\pm20$ meV, and apical O at 60$\pm15$ meV \cite{thom,coh,chen}.
Our estimations of $N_{site}$ from ref. \cite{chen} are shown in the last part of Table I. 
A rescaling is made for the wavelength. The LMTO results are calculated for short waves, 8$a_0$, but
the waves are longer at the doping $x=0.16$ (corresponding to $q \approx 0.08$), when
 $V_q^p$ and $V_q^m$ are larger
by factors of $\sim 1.1$ and $\sim 1.5$, respectively \cite{tj6}.
These factors  might be fine tuned later if the calculated wave lengths fall outside
the expected range, but in
the first calculations we calculate the total $V_q^t$ from,
\begin{equation}
V_q^t = \sum_{i} (1.1 * V_{q,i}^p + 1.5 * V_{q,i}^m) * N_i
\end{equation}
where $i$ is the site index. The resulting $V_q^t$ are 17, 18, 23 and 22 mRy at the energies centered around
15 (La), 25 (Cu), 50 (plane-O) and 60 meV (apical-O), respectively. These values are used in the subsequent calculations of
the $q-\omega$ dispersion.
As discussed later, spin waves at higher energy are independent of the phonons, and $V_q^p = 0$. 

\begin{table}[b]
\caption{\label{table1} Three first lines: Phonon distortion amplitudes, $u/a_0$, 
induced potential shifts, $V_q^p$ (mRy), and exchange splitting, $V_q^m$ (mRy), caused
by spinwaves ($\mu_BH = \pm 5$ mRy). 
All shifts/splittings refer to the maximal value on a Cu-site,
and they are determined in self-consistent
LMTO calculations for supercells of length $8a_0$.
Remaining lines: The partial characters of the phonon DOS, N$_i$, for different sites, $i$,
as estimated from ref \cite{chen}.
 }
\vskip 5mm
\begin{center}
\begin{tabular}{l c c c c c }
\hline
 wave & no-phon & pl-O$_x$ & La$_z$ & ap-O$_z$ & Cu$_x$\\
\hline \hline
$u/a_0$ & - & 0.014& 0.021& 0.017& 0.024 \\
$V_q^p$  & - &  15 & 2 & 5 & 3.5 \\
$V_q^m$ & 8 & 12.5 & 12  & 10 & 8.5  \\
\hline
N$_{pl-O}$ & - & 0.6 & 0.0 &  0.2 & 0.2 \\
N$_{La}$   & - & 0.0 & 0.65 &  0.0 & 0.35 \\
N$_{ap-O}$ & - & 0.35 & 0.0 &  0.65 & 0.0 \\
N$_{Cu}$   & - & 0.2 & 0.2 &  0.0 & 0.6 \\
\hline
\end{tabular}
\end{center}
\end{table}

In a second step we use the parameters $V_q^t$ in
a 2D nearly free-electron (NFE) model.
The AFM spin arrangement on neighboring Cu along [1,0,0] in
undoped LBCO corresponds to
a potential perturbation, 
$V(\bar{x}) = V_q^t exp(-i\bar{Q} \cdot \bar{x})$ (and equivalently with $\bar{y}$ along [0,1,0]). The periodicity
in real space is defined through the Cu-Cu distance, and
a gap of size 2$V_q^t$ appears at the zone boundary, at $\bar{Q}/2$. 
A further modulation ($\bar{q}$) of this order into 1D-stripes perpendicular to $\bar{x}$
(or "checkerboards" in 2D along $\bar{x}$ and $\bar{y}$) is achieved by a multiplication of the potential
by $exp(i\bar{q} \cdot \bar{x})$, where $\bar{q} < \bar{Q}$. Totally,
this makes
$V(\bar{x}) 
= V_q^t exp(-i\bar{Q}_x \cdot \bar{x})$, where $\bar{Q}_x =\bar{Q}-\bar{q}$. 
The periodicity of this potential is now larger, and
the gap moves away from $\bar{Q}/2$ to ($\bar{Q}-\bar{q}$)/2.
Magnetic side spots appear at $\bar{q}/2$ surrounding $\bar{Q}/2$ in
probes which can separate the two periodicities.
A 3x3 eigenvalue problem with matrix elements 
$H_{11}=E-k_x^2-k_y^2$, $H_{22} =E-(k_x-Q_x)^2-k_y^2$, $H_{33}=E-k_x^2-(k_y-Q_y)^2$,
$H_{12}=H_{13}=V_q^t$ and $H_{23}=0$, is solved. 
 The lowest FE
band (when $V_q^t=0$) contains 2 electrons up to $E_F \sim 0.15 Ry$, when the effective mass is one.
The total band width is quite
close to the real band.
The model is entirely 2-D, (the DOS is constructed from a sum over all states
in the plane), which is reasonable in view of the very small band dispersion along $\bar{k_z}$ of
real band structures for the cuprates \cite{tj1}.

An important result of the
2D-NFE model is that it leads to a correlation between doping and the amplitude of $V_q^t$ \cite{tj6}.
The reason is that the gap opens along $(k_x,0)$ and $(0,k_y)$, but not
in the diagonal direction.
The combined effect is that the dip in the total DOS
(at which $E_F$ should fall for optimal doping)
will not appear at the same band filling for a small and a wide gap, even if the q-vector is the same. 
Alternatively, since the gap should appear at the same energy (at $E_F$) for different $V_q^t$, one has to vary
the q-vector until it fits. The present calculations are made for 0.16 holes per Cu (which is close
to x in ref. \cite{vig}) with $V_q^t$ from eq. 1.

The results are shown in fig. 1, with positive and negative $q$ 
displayed symetrically, and the spectrum is shaped like
an hour-glass with a "waist" at intermediate energy. 
The points below 70 meV are for the coupling to
the 4 types of phonons.
The general shape of the $q-\omega$ dependence and the amplitudes of $q$ are similar as in ref. \cite{vig}, 
with the smallest
$q$ at about 0.05 for SPC dominated by O-p phonons around 50 meV. The weaker SPC
for La at low energy makes $q$ larger. The calculated results depend on the different
parameters in the NFE model. Larger band mass leads to smaller amplitudes of $q$. The
variations of $q$ is a result of the differences in $V_q^t$ at different energies.
Consequently, if  no mixing of the phonon modes (through the $N_{site}$-coefficients) were made 
it would increase
the variations of $q$,
even though the general hour-glass shape would remain.

The uppermost part of the spectrum needs some comments. 
Spin waves and phonons are tied together at the same frequency and $q$-vector for the
optimal mechanism of SPC. 
Spin waves with higher frequency than the phonons cannot profit from this mechanism.
Thus, it is assumed that spin waves with frequencies higher than the phonon frequencies
are independent of lattice vibrations, whereby $V_q^p = 0$. 
This "phonon independent" result is put rather arbitrarily at $\hbar\omega \sim$120 meV
in fig. 1,
which is about twice the highest phonon frequency. 
The calculated wavelengths for the SPC modes below $\sim$70 meV are found to be longer
than the length of the supercell in the LMTO
calculation, which justifies the factors 1.1 and 1.5 in eq. 1. 
However, the solution without SPC 
leads to a wave which
is even shorter than the cell in the LMTO
calculation. The the factor 1.5, in eq. 1, is not justified for this (phonon free) mode, and
a corrected calculation with $V_q^t =$8 mRy is more appropriate,
with a solution for $q=0.24$, see fig. 1.
In contrast to the cases with SPC and larger $V_q^t$, this solution 
implies a very short wave, 4-5 $a_0$ only, where
$V_q^t$ might be reduced even more. This will, because of the self-consistent feed-back between spin density and potential,
lead towards a vanishing spin wave \cite{tj5}. Therefore, it is difficult to imagine larger $q$ for spin
excitations with small $V_q^t$ at this doping. 
 
Other high-energy solutions with small $V_q^t$ exist for non-equal $q$-vectors
along $x$ and $y$ \cite{tj6}, where the upper of two gaps corresponds to a doping of $0.16$.
These solutions are found within some
range of $q$, but the dips in the DOS are relatively weak. A satisfactory
solution is found for $q_x$ and $q_y$ near 0.14 and 0.11, respectively. The average of the two vectors, 0.125,
is comparable to the wave length in the LMTO calculation (where $V_q^m$=8 mRy).
Two gaps remain in the DOS for larger separation
of the two $q$-vectors, but the dips are found at too high and too low doping.
The existence of this multitude of solutions indicates that 
large broadening and damping is expected at high energies.
Such fluctuations are  spread in energy and momentum. They decay rapidly,
 since they are not linked to the phonon spectrum.  
 
Less doping will make the waves longer
and the waist in the $\bar{q}-\hbar\omega$ diagram becomes narrower.
The narrowing should be noticed at all energies  even at the highest E where spinwaves are
excited without the help of phonons. Smaller $q$-vectors have been observed recently
in lightly doped La$_{1.96}$Sr$_{0.04}$CuO$_4$ \cite{mat}, in agreement
with this prediction. However, the spin modulation has turned from parallel
(the Cu-O bond) to diagonal direction at such low doping. In addition, the waist in the observed 
spectrum is found at lower energy than for parallel doping \cite{mat},
which in the SPC-model implies increased SPC for phonons at lower energy. Ab-initio calculations for 
modulations along the
diagonal direction are needed to verify if La- or Cu-modes grow in importance relative
to the high energy O-modes.

Heavier O-isotopes will decrease the 
frequencies for the phonons and the coupled spin waves,
and move the waist to lower E.
They may also decrease $u$, at least at 
low temperature, which in the model
for SPC will make $V_q^t$ smaller. From the correlation between $V_q^m$ and wave length
it is expected that $q$ becomes larger. In all, this leads to a wider waist at
lower energy, while the upper part of the spectrum should be insensitive to isotope
shifts. 

\begin{figure}
\includegraphics[height=7.0cm,width=8.0cm]{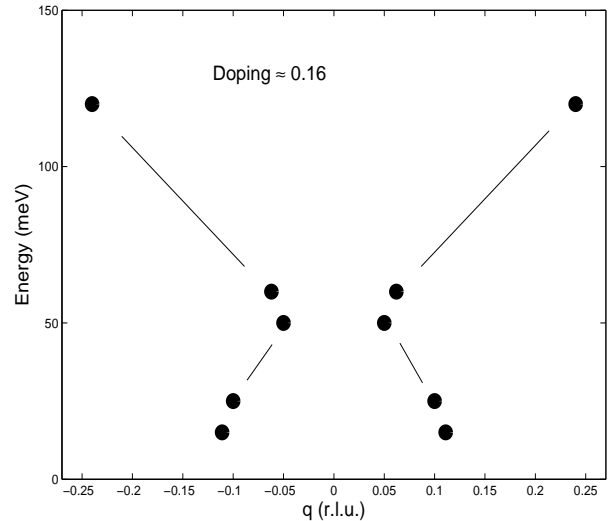}
\caption{Calculated $q-\hbar\omega$ relation from the 2D-NFE model and
the parameters $V_q^t$ for doping
$x=$0.16 as in the experiment of Vignolle {\it et al} \cite{vig}. The lines connecting the points indicate
the expected continuation for intermediate energies. The solution without SPC, the uppermost point,
is not very precise (see text).
}
\label{fig1}
\end{figure}

Undoped cuprates are generally stable AFM insulators.
The exchange enhancement for spin waves in doped cuprates is moderately
large in the absense of phonons, and it becomes
peaked if a phonon can intervene. This can be translated into a large
phonon deformation potential. The unusual part of it is that this deformation
potential resides in the equal spin part, and hence it
is more correct to associate this deformation potential with a large $\lambda_{sf}$
for spin fluctuations.  
The calculated $V_q$-parameters determine the spin excitation spectrum for
large SPC, but they are also the key parameters for estimates of superconducting $T_c$.
For instance, $V_q^p$ corresponds
to the monopolar matrix element for electron-phonon coupling, $\lambda_{ph}$.
The good agreement between the calculated spin excitations and experiment
suggests that the calculated $V_q$'s are of the correct order.
It is not clear how the contribution to $T_c$ from
coexisting phonon and spin fluctuations will be shared, but a
simple account for separate $\lambda_{ph}$ 
 and $\lambda_{sf}$ gives rather large
values for $T_c$. The estimates are based on $\lambda = {\it N} I^2/K_u$,
where ${\it N}$ is the DOS at $E_F$ ($\approx 0.5 (eV \cdot Cu \cdot spin)^{-1}$), and
$I= \langle \Delta V^p/\Delta u \rangle$.
From the Table and with $a_0 = 3.78 \AA$, $\Delta V^p/\Delta u \sim 3.8 eV/\AA$
for a pure (plane) O-phonon, which makes $\lambda_{ph} \approx 0.5$, for $K_u$ of the order 15 $eV/\AA^2$.
If phonons stir up a spin fluctuation (through SPC) one should only
consider the equal spin part in $\Delta V^m$, which is almost of the same order
as $\Delta V^p$ for the O-phonon (see Table). 
Moreover, one should add a $K_m$
to $K_u$ because of the energy of the spin wave, but this contribution
can be negative if the coexistence makes the phonon softer ($K_m$ without phonon may be large and positive).
By assuming $K_m \sim 0$ this gives  
$\lambda_{sf} \approx $ 0.4, and it will increase faster than $\lambda_{ph}$ when the doping becomes smaller.
The BCS formula $T_c = 1.13 \cdot \hbar \omega \cdot exp({-1/\lambda})$
 makes $T_c$ of the order 80-40 K, for these two $\lambda$'s,
when $\hbar \omega = 50 meV$.
This is very approximate, but it shows that the coupling strength can
be sufficient for a large $T_c$. Note also that since the exchange enhancement
in LSDA is too weak to stabilize AFM in undoped LSCO, it will also
produce a too low value of $\lambda_{sf}$.

The total SPC becomes too large at underdoping when the wave modulations are longest, so $N$
(and $\lambda$) are reduced because of the pseudogap, and hence the pseudogap competes
with superconductivity. On the overdoped side $\lambda$
is reduced because of a small $\Delta V$, and $T_c \rightarrow 0$. 
Optimal conditions for a high $T_c$ are found at intermediate doping, but
the estimates are not precise enough for finding the exact value or the exact doping limits for
vanishing $T_c$. The role of phonons decreases as $x \rightarrow 0$, since the ratio $V^m/V^p$ is growing 
at small doping.

In conclusion, these calculations suggest that the hour-glass shaped dispersion of the spin wave
spectrum is a consequence of different degrees of SPC for different phonon modes.
The "half-breathing" O-mode in the mid-part of the phonon spectrum is found to be the most efficient one,
which explains
the waist in the spin wave dispersion.
The SPC-NFE model simulates several normal state
properties surprisingly well \cite{tj6}.
There is nothing special with the electronic band structure in
this scenario. The LDA bands are essentially correct for doped systems, although
the exchange enhancement is underestimated
in normal LSDA calculations \cite{tj4}.
Finally, it is argued that the observed spin excitations provide
important information about the mechanism of superconductivity, since
the excitation spectrum can be related directly to the coupling parameters.
In particular, there is an unusual contribution to the equal spin pairing 
parameter $\lambda_{sf}$
coming from the coupling to phonons.

I am grateful to B. Barbiellini and C. Berthod for various discussions.

\end{document}